\newcommand{\CLASSINPUTtoptextmargin}{1.8cm}
\newcommand{\CLASSINPUTbottomtextmargin}{4.2cm}
\documentclass[conference,a4paper]{IEEEtran}

\usepackage{algorithm,algorithmic}
\usepackage{amsmath,amssymb,mathrsfs,amsbsy}
\usepackage{cite}
\usepackage{subfigure}
\usepackage{graphicx,color}

\usepackage{hyperref}
\usepackage{subfigure}
\usepackage{balance}
\usepackage{tikz,siunitx}
\usepackage{url}

\newtheorem{lem}{Lemma}

\newtheorem{thm}{Theorem}

\definecolor{sblue}{RGB}{0,51,120}
\definecolor{sred}{RGB}{200,51,130}

\newcommand{\matc}[1]{\mbox{\boldmath $\mathcal{#1}$}}

\ifCLASSINFOpdf
\else
\fi

\begin{document}

\title{Constellation-Oriented Perturbation for Scalable-Complexity MIMO Nonlinear Precoding}

\author{Jinfei Wang$^{1}$, Yi Ma$^{1\dagger}$, Na Yi$^{1}$, Rahim Tafazolli$^{1}$, and Fei Tong$^{2}$\\
	{\small $^{1}$5GIC and 6GIC, Institute for Communication Systems, University of Surrey, Guildford, UK, GU2 7XH}\\
		{\small $^{1}$Emails: (jinfei.wang, y.ma, n.yi, r.tafazolli)@surrey.ac.uk}\\
		{\small $^{2}$Samsung Cambridge Solution Centre, Email: f.tong@samsung.com}
		}
\markboth{}%
{}

\maketitle

\begin{abstract}
In this paper, a novel nonlinear precoding (NLP) technique, namely constellation-oriented perturbation (COP), is proposed to tackle the scalability problem inherent in conventional NLP techniques.
The basic concept of COP is to apply vector perturbation (VP) in the constellation domain instead of symbol domain; as often used in conventional techniques.
By this means, the computational complexity of COP is made independent to the size of multi-antenna (i.e., MIMO) networks.
Instead, it is related to the size of symbol constellation. 
Through widely linear transform, it is shown that COP has its complexity flexibly scalable in the constellation domain to achieve a good complexity-performance tradeoff. 
Our computer simulations show that COP can offer very comparable performance with the optimum VP in small MIMO systems.
Moreover, it significantly outperforms current sub-optimum VP approaches (such as degree-$2$ VP) in large MIMO whilst maintaining much lower computational complexity.
\end{abstract}

\section{Introduction}\label{secI}
In the scope of signal processing for communications, one of fundamental problems lies in the scalability of nonlinear precoding (NLP) techniques for wireless multi-antenna (i.e., MIMO) downlink transmissions. 
This problem is particularly important for large MIMO system in ill-conditioned channels, where low-complexity linear precoding techniques such as matched filter and (regularized) zero forcing (R)ZF are too sub-optimum \cite{4599181}. 
On the other hand, advanced NLP techniques such as vector perturbation (VP) \cite{1413598} and Tomlinson-Harashima precoding (THP) \cite{6151847} have their computational complexities scaling exponentially with the size of MIMO networks.
A wise way of mitigating the scalability problem is to adopt the massive-MIMO \cite{5595728} or asymmetric MIMO architecture \cite{8354786}. 
When wireless channels are wide-sense stationary uncorrelated scattering (WSSUS), those highly over-determined MIMO systems are well conditioned and linear precoding techniques become near-optimum \cite{Rusek6375940}. 

With ever growing demand for the spectral efficiency, user terminals (not only smart phones but also many other mobile computing devices) will be equipped with more antennas. 
This is pushing multiuser-MIMO back to the symmetric architecture. 
Moreover, wireless channels can be spatially more correlated.
It is possible to add more service antennas at transmitter in order to maintain the service-to-user antenna ratio. 
However, such will significantly increase the hardware cost and set a challenging task to device manufactory for the calibration of antenna-array \cite{1350238}. 
Moreover, when the operating frequency is fixed, adding more antennas often means enlarging the aperture of antenna-array.  
This could drag user terminals into the near field of the transmitter and renders wireless channels spatially non-stationary \cite{BJORNSON20193n,Wang2022}.
Also, as discussed in \cite{9685536}, current MIMO transceivers are not yet optimized for non-stationary MIMO channels. 
As a conclusion of our brief background review, (close-to) symmetric MIMO architecture is still one of important candidates for future wireless MIMO technology, where scalable NLP technique plays a central role. 

\begin{figure}[t]
\centering
\includegraphics[scale=0.18]{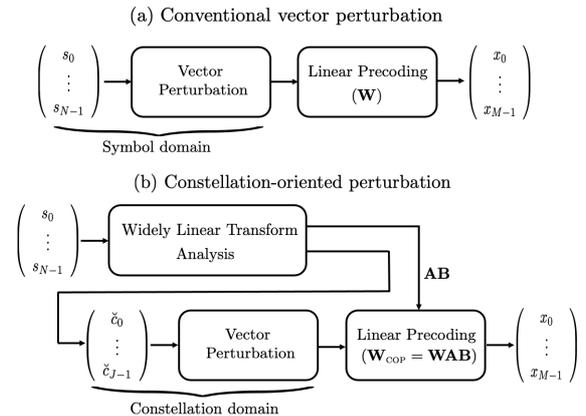}
\caption{Illustration of the major difference between the constellation-oriented perturbation and the conventional vector perturbation.}
\vspace{-1em}
\label{fig1}
\end{figure}

In this paper, we propose a novel NLP technique named constellation-oriented perturbation (COP). 
The block diagram of COP is illustrated in Fig. \ref{fig1}-(b).
Different from the conventional VP technique that applies vector perturbation in the symbol domain (see Fig. \ref{fig1}-(a) and Eqn. \eqref{eqn05}), COP applies perturbation in the constellation domain (also refer to Eqn. \eqref{eqn10}). 
By this means, the computational complexity of COP is made independent to the size of MIMO networks. 
Instead, it is only related to the size of symbol constellation (or equivalently the modulation order).
Moreover, it is shown that COP can flexibly scale its computational complexity in the constellation domain through widely linear transform (see {\it Theorem \ref{thm01}}). 
This is appealing in the sense that COP is scalable with the modulation order and able to achieve the best complexity-performance tradeoff. 

In addition to the novel concept, the complexity and optimality of COP are also carefully studied and compared with current state-of-the-art (see Section \ref{sec04} for the detailed discussion of baseline selection).
It is revealed that COP is a sub-optimum NLP technique when comparing to the optimum VP \cite{1413598}. 
Nevertheless, it offers very comparable performance to the optimum VP (around $1\sim 3$ dB in $E_\mathrm{s}/N_0$) in small MIMO systems (e.g., $8$-by-$8$ MIMO).
With the increase of MIMO size, the optimum VP can be hardly implemented due to its prohibitive complexity. 
This motivates the performance comparison between COP and the degree-K VP (DKVP) approach reported in \cite{7397882}.
It is demonstrated that COP can significantly outperform D2VP (i.e., $K=2$ for DKVP) particularly for large sizes of MIMO (e.g., $10\sim20$ dB gain for $512$-by-$512$ MIMO).
Moreover, COP offers much lower complexity than the D2VP approach.
All of the above observations are elaborated by Monte Carlo simulations.

\section{Preliminaries and Problem Statement}
\subsection{NLP for MIMO Downlink Transmission}
Consider a multiuser-MIMO downlink system where an access point with $M$ transmit-antennas serves a set of user terminals ($N$). 
To facilitate our technical presentation, it is assumed that each user terminal has only a single receive-antenna.
Denote $\mathbf{s}\triangleq[s_0,...,s_{N-1}]^T$ to be the information-bearing symbol block, where $s_n$ stands for the symbol sent to the $n^{th}$ user terminal and $[\cdot]^T$ for the vector/matrix transpose.
Prior to transmission, the symbol block $(\mathbf{s})$ is fed into a precoder $\Gamma$: $\mathbb{C}^{N}\rightarrow\mathbb{C}^{M}$ to yield a block $\mathbf{x}=\Gamma(\mathbf{s})$.
Then, the precoded block $(\mathbf{x})$ goes through the MIMO channel $(\mathbf{H}\in\mathbb{C}^{M\times N})$ and arrives at the $n^{th}$ user terminal in the following signal format
\begin{equation}\label{eqn01}
y_n=\mathbf{h}_n^T\mathbf{x}+v_n,~n=0,...,N-1,
\end{equation}
where $y_n$ is the received noisy symbol, $\mathbf{h}_n^T$ is the $n^{th}$ row of $\mathbf{H}$, and $v_n\sim CN(0, \sigma^2)$ is the thermal noise. 
Finally, each receiver employs a demodulator $\Omega: \mathbb{C}^{1}\rightarrow\mathbb{C}^{1}$ to reconstruct the original symbol: $\hat{x}_n=\Omega(y_n)$.

The fundamental aim of the precoding research is to find a precoder $\Gamma(\cdot)$ and its corresponding decoder/demodulator $\Omega(\cdot)$ which can satisfy various requirements set to the multiuser-MIMO downlink communication.
As already reviewed in Section I, there have been already many contributions within this scope.
Specifically for the VP technique originally proposed in \cite{1413598}, the precoder can be expressed by
\begin{equation}\label{eqn02}
\mathbf{x}=\mathbf{W}(\mathbf{s}+\alpha\mathbf{b}),
\end{equation}
where $\mathbf{W}\in\mathbb{C}^{M\times N}$ is a linear precoding matrix, $\alpha$ is a scaling factor (often set $\alpha>2$) known to both the transmitter and receivers, and $\mathbf{b}$ is the perturbation vector, with each element of which drawn from a finite set of complex integers $(\mathcal{A})$.
For the simplest case (ZF precoding) when $\mathbf{W}=\mathbf{H}^H(\mathbf{H}\mathbf{H}^H)^{-1}$, the linear model \eqref{eqn01} becomes
\begin{equation}\label{eqn03}
y_n=s_n+\alpha b_n+v_n,~n=0,...,N-1,
\end{equation}
where $b_n$ is the $n^{th}$ element of $\mathbf{b}$. 
Then, the following modulo receiver can be employed to reconstruct the original symbol
\begin{equation}\label{eqn04}
\hat{s}_n=y_n-\alpha\left\lfloor\frac{y_n}{\alpha}+\frac{1}{2}\right\rfloor,
\end{equation}
where $\lfloor\cdot\rfloor$ stands for the integer floor.

\subsection{Research Problem and Motivation}\label{sec2b}
Despite appealing in its concept, VP must pay computational complexity of $\mathcal{O}(|\mathcal{A}|^N)$ in order to find the optimum perturbation vector $\mathbf{b}^\star$ via
\begin{equation}\label{eqn05}
\mathbf{b}^{\star}=\underset{\mathbf{b}}{\arg\min}\|\mathbf{W}(\mathbf{s}+\alpha\mathbf{b})\|^2,
\end{equation}
where $\|\cdot\|$ is the Euclidean norm, and $|\mathcal{A}|$ is the cardinality of $\mathcal{A}$. 
Therefore, the original VP technique is not scalable to the size of the MIMO network $(N)$. 
To tackle the challenge of scalability, it has been recently proposed to apply perturbation only on several symbols ($K$) of $\mathbf{s}$ \cite{7397882}.
Such trades off the performance optimality for reduced computational complexity $\mathcal{O}({N \choose K}|\mathcal{A}|^K$).
Sometimes, this complexity could be still too high even for the case of D2VP (i.e., $K=2$). 
Our question is: would it be possible to find a low-complexity NLP approach which can achieve better performance-complexity tradeoff than current approaches? 
This forms the motivation of the COP technique presented in Section \ref{secIII}.

\section{The Constellation-Level Perturbation Approach}\label{secIII}
\subsection{The Basic Concept}
In a nutshell, the idea of COP is to apply the perturbation operation on a so-called constellation level instead of symbol level in the conventional sense. 
By such means, we will show that a better performance-complexity tradeoff can be achieved particularly for wireless MIMO systems with higher-order modulations.
To convey the concept in an easy-to-access way, we start with a revisit of how the symbol block $(\mathbf{s})$ is formed.  

It is understood that each element of $\mathbf{s}$ is drawn from a finite-alphabet set: $s_n\in\mathcal{C}\triangleq\{c_l|l=0,...,L-1\}$.
Then, we form a vector $\mathbf{c}=[c_0,...,c_{L-1}]^T$ by collecting all elements inside $\mathcal{C}$. 
A mapping matrix $\mathbf{A}\in\mathbb{R}^{N\times L}$ is formed to map $\mathbf{c}$ onto $\mathbf{s}$:
\begin{equation}\label{eqn06}
\mathbf{s}=\mathbf{A}\mathbf{c},
\end{equation}
where each row of $\mathbf{A}$ has only a single $'1'$ element indicating the place where the corresponding symbol is mapped, and all other elements of $\mathbf{A}$ are zeros.
It is possible that $\mathbf{A}$ can have a column with all elements to be zero, in which the size of $\mathbf{A}$ and $\mathbf{c}$ can be reduced.

The concept of COP is to form the transmitted block $(\mathbf{x})$ in the following way:
\begin{equation}\label{eqn07}
\mathbf{x}=\mathbf{W}\mathbf{A}(\mathbf{c}+\tau\matc{\rho}),
\end{equation}
where the parameter pair ($\tau$, $\matc{\rho}$) plays the same role as ($\alpha$, $\mathbf{b}$) in \eqref{eqn02};
we use different notations to avoid any possible confusion, and $\tau$ can have a different optimum value from $\alpha$ (shown in Section IV). 
With \eqref{eqn07}, the optimization problem becomes
\begin{equation}\label{eqn08}
\matc{\rho}^{\star}=\underset{\matc{\rho}}{\arg\min}\|\mathbf{W}\mathbf{A}(\mathbf{c}+\tau\matc{\rho})\|^2.
\end{equation}
Then, the computational complexity becomes $\mathcal{O}(|\mathcal{A}|^L)$, which is no longer dependent on the size of MIMO network $(N)$.
The complexity is considerably small for lower-order modulations such as $4$QAM ($L=4$).
One might argue that the proposed approach is not scalable to the modulation order $(L)$. 
It is indeed true at least for the mathematical form presented in \eqref{eqn07}-\eqref{eqn08}. 
Therefore, a widely linear form of COP is proposed in Section \ref{SecIIIb}. 

\subsection{The WL-COP Approach with Scalable Complexity}\label{SecIIIb}
Different from the standard WL form in other papers (e.g., \cite{7331280,7882699}), we are only interested in the WL form of the constellation vector ($\mathbf{c}$). 
\begin{lem}\label{lem01}
Assume: {\it C1)} $L$ to be a square number; 
{\it C2)} symmetric constellation of modulation symbols, i.e., $\Re(c_l)$ and $\Im(c_l)$ share the same finite-alphabet set ($\mathcal{B}$) with the size $J=\sqrt{L}$.
Form a vector $\breve{\mathbf{c}}\in\mathbb{R}^{J}$ by collecting all elements from $\mathcal{B}$.
Then, $\mathbf{c}$ and $\breve{\mathbf{c}}$ can be related by a mapping matrix $\mathbf{B}\in\mathbb{C}^{L\times J}$:
\begin{equation}\label{eqn09}
\mathbf{c}=\mathbf{B}\breve{\mathbf{c}},
\end{equation}
\end{lem}
where each row of $\mathbf{B}$ must have a single $'1'$ element and a single $'j'$ element. 

The proof of {\em Lemma  \ref{lem01}} is straightforward and thus omitted.
Applying \eqref{eqn09} into \eqref{eqn08} yields 
\begin{equation}\label{eqn10}
\breve{\matc{\rho}}^{\star}=\underset{\breve{\matc{\rho}}}{\arg\min}\|\mathbf{W}\mathbf{AB}(\breve{\mathbf{c}}+\tau\breve{\matc{\rho}})\|^2,
\end{equation}
where the perturbation vector ($\breve{\matc{\rho}}$) is from a real set.  
By this means, the computational complexity reduces to $\mathcal{O}(\sqrt{|\mathcal{A}|}^{J})$. 
This complexity is already quite low for some higher-order modulations such as $64$QAM ($J=8$).
Nevertheless, we can further reduce the complexity by manipulating the mapping matrix $\mathbf{B}$ and the vector $\breve{\mathbf{c}}$.
\begin{thm}\label{thm01}
Given the conditions in {\em Lemma \ref{lem01}}, there exists a sparse matrix $\mathbf{B}\in\mathbb{C}^{L\times J}$
which forms the relationship \eqref{eqn09}. 
Each row of $\mathbf{B}$ has two non-zero elements (denoted by $b_n^{(1)}$ and $b_n^{(2)}j$, respectively), 
which can be used to scale the size $J$ within the range of $J\in[1, \sqrt{L}]$.
\end{thm}

\begin{proof}
The intuition of {\it Theorem \ref{thm01}} is that the complexity of WL-COP can be managed by means of forming the mapping matrix $\mathbf{B}$.
Taking $256$QAM as an example ($L=256$), we can set $b_n^{(1)}, b_n^{(2)}\in\{1\}$ on each row of $\mathbf{B}$. 
This is the case which has been addressed in {\it Lemma \ref{lem01}}; we have $J=\sqrt{L}=16$.
However, the complexity is still too high for practical uses and it must be further reduced. 
To this end, we can set $b_n^{(1)}, b_n^{(2)}\in\{1, -1\}$. 
In this case, the vector $\breve{\mathbf{c}}$ can have all negative (or positive) numbers removed; 
such halves the size $J$ (i.e., $J=8$).
Accordingly, we can further reduce the size ($J$) by growing the finite-alphabet set for $b_n^{(1)}, b_n^{(2)}$ until the case $J=1$. 
Hence, the complexity of WL-COP is scalable with the modulation order. 
\end{proof}

\subsection{The Optimality Study}\label{SecIIIc}
To facilitate our study on the optimality of WL-COP, it would be important to firstly understand the principle of the original VP technique as well as its difference from the principle of WL-COP. 
\subsubsection{Principle of VP and DKVP}
We start from the special case stated in {\it Lemma \ref{lem02}}.
\begin{lem}\label{lem02}
Given the conditions: {\it C3)}  $\mathbf{W}^H\mathbf{W}=\mathbf{\Lambda}$ (diagonal matrix); 
{\it C4)} $\alpha>2\max|s_n|$ (condition to enable the modulo receiver), the solution to \eqref{eqn05} is: $\mathbf{b}=\mathbf{0}$.
\end{lem}
\begin{proof}
Given the condition {\it C3)}, the Euclidean norm in \eqref{eqn05} can be rewritten into
\begin{IEEEeqnarray}{ll}\label{eqn11}
\|\mathbf{W}(\mathbf{s}+\alpha\mathbf{b})\|^2&=\sum_{n=0}^{N-1}\lambda_n|s_n+\alpha b_n|^2\\
&\geq\sum_{n=0}^{N-1}\lambda_n|s_n|^2,\label{eqn12}
\end{IEEEeqnarray}
where $\lambda_n(>0)$ is the diagonal element of $\mathbf{\Lambda}$. 
The inequality in \eqref{eqn12} is due to the condition {\it C4)}, and the equality holds only when $b_n=0, \forall n$. 
{\it Lemma \ref{lem02}} is therefore proved.
\end{proof}

{\it Lemma \ref{lem02}} has two implications: 
{\it 1)} VP has no gain when the linear precoding matrix $\mathbf{W}$ satisfies the condition of vector-wise orthogonality;
{\it 2)} The perturbation vector $(\mathbf{b})$ dominates the effectiveness of VP when vector-wise correlation exists. 

When $\mathbf{W}$ does not satisfy the condition {\it C3)}, VP searches for the optimum solution over a finite set ($\Phi_{\textsc{vp}}$) which has $|\mathcal{A}|^{N}$ candidates. 
For the DKVP approach introduced in Section \ref{sec2b}, it searches for the best solution over a finite set ($\Phi_{\textsc{dkvp}}\in\Phi_{\textsc{vp}}$), which has ${N \choose K}|\mathcal{A}|^K$ candidates.
In terms of the optimality of DKVP, the following result has been given in \cite{7397882}:
\begin{lem}\label{lem03}
For a sufficiently large $K$ (e.g., $20\%\sim25\%$ of $N$), the optimum VP solution ($\mathbf{b}^\star$) falls into the finite set of DKVP, i.e., $\mathbf{b}^\star\in\Phi_{\textsc{dkvp}}$.
\end{lem}

In light of the above discussion, we rewrite \eqref{eqn02} into  
\begin{equation}\label{eqn13}
\mathbf{x}=\underbrace{\mathbf{W}_1\mathbf{s}_1}_{\triangleq\mathbf{x}_1}+\mathbf{W}_2\underbrace{(\mathbf{s}_2+\alpha\mathbf{b}_2)}_{\triangleq\bar{\mathbf{b}}},
\end{equation}
where $\mathbf{b}_2\in\mathbb{C}^{\overline{N}}$ is formed by all non-zero elements ($\overline{N}$) of $\mathbf{b}$;
$\mathbf{s}_2\in\mathbb{C}^{\overline{N}}$ is formed by $\overline{N}$ elements of $\mathbf{s}$ corresponding to $\mathbf{b}_2$;
$\mathbf{W}_2\in\mathbb{C}^{M\times\overline{N}}$ is formed by $\overline{N}$ column vector of $\mathbf{\mathbf{W}}$ corresponding to $\mathbf{b}_2$;
$\mathbf{W}_1$ and $\mathbf{s}_1$ are formed by the rest part of $\mathbf{W}$ and $\mathbf{s}$, respectively.
Then, the VP optimization problem \eqref{eqn05} becomes
\begin{equation}\label{eqn14}
\bar{\mathbf{b}}^{\star}=\underset{\bar{\mathbf{b}}}{\arg\min}\|\mathbf{x}_1+\mathbf{W}_2\bar{\mathbf{b}}\|^2.
\end{equation}
According to {\it Lemma \ref{lem02}}, $\mathbf{x}_1$ must be correlated with every column vector of $\mathbf{W}_2$; 
or otherwise, $\mathbf{b}_2$ would contain zeros. 
Moreover, the performance of VP is directly proportional to the correlation between $\mathbf{x}_1$ and $\mathbf{W}_2$.  

Specifically for the DKVP approach, it can achieve the optimum VP solution when $K\geq\overline{N}$. 
Given that $\overline{N}$ increases linearly with $N$, DKVP must have its complexity increasing exponentially with $N$ in order to preserve the optimality.
Practically, due to the complexity constraint, the parameter $K$ cannot be made large (often $K=2\sim4$).
This is however too small for a large MIMO, where we will easily have the condition of $N\gg K$. 
In this case, $\mathbf{W}_2$ is only a very small portion of $\mathbf{W}$. 
Such renders the correlation between $\mathbf{x}_1$ and $\mathbf{W}_2$ very weak and consequently makes the DKVP approach too sub-optimum.

\subsubsection{Optimality of WL-COP}
Mathematically, \eqref{eqn10} shows that WL-COP differs from VP mainly in its linear precoding matrix denoted by
\begin{equation}\label{eqn15}
\mathbf{W}_{\textsc{cop}}=\mathbf{WAB}.
\end{equation}
When $\mathbf{W}$ satisfies the condition {\it C3)}, $(\mathbf{W}_{\textsc{cop}}^H\mathbf{W}_{\textsc{cop}})$ is also a diagonal matrix because the role of $(\mathbf{AB})$ is mainly to divide columns of $\mathbf{W}$ into several groups. 
In this case, WL-COP also gives: $\breve{\matc{\rho}}=\mathbf{0}$.

When $\mathbf{W}$ does not satisfy the condition {\it C3)}, the objective function \eqref{eqn10} cannot be represented into a similar form as \eqref{eqn14}.
This is because $\mathbf{W}$ is divided into $J$ groups; 
column vectors within each group are linearly combined. 
Given that $J$ is considerably smaller than $N$, each column of $\mathbf{W}_\textsc{cop}$ is a linear combination of many column vectors (averagely $(N)/(J)$) of $\mathbf{W}$. 
This makes columns of $\mathbf{W}_\textsc{cop}$ highly correlated with each other. 
In this case, similar result as {\it Lemma \ref{lem03}} does not exist for WL-COP.
As a consequence, WL-COP can only offer a sub-optimum VP solution.
However, the performance of WL-COP would not be as sub-optimum as DKVP because all column vectors of $\mathbf{W}_\textsc{cop}$ are highly correlated. 
Section IV shows that WL-COP can offer very comparable performance to the optimum VP at least in small MIMO systems \footnote{It is hard to evaluate the performance gap between WL-COP and the optimum VP in large MIMO, where the optimum VP is computationally prohibitive.}.
Moreover, WL-COP significantly outperforms D2VP in large MIMO systems whilst maintaining much lower computational complexity. 

\section{Simulation Results and Discussion}\label{sec04}
Computer simulations were carried out to evaluate the COP/WL-COP technique presented in Section \ref{secIII}.
For the sake of conciseness, we use COP as the unified terminology for both as they have no fundamental difference.
As far as the spectral efficiency and MIMO scalability are concerned, we choose higher-order modulations (mainly $64$, $256$, and $1024$QAM) and larger MIMO sizes (mainly $128$-by-$128$ and $512$-by-$512$ MIMO).
The performance metric is chosen to be symbol-error-rate (SER) averaged over sufficient Monte Carlo trials.
For each trial,  the wireless MIMO narrowband channel was generated according to independent complex Gaussian distribution (Rayleigh in amplitude); this is the commonly used simulation setup in the literature.
The signal-to-noise ratio (SNR) is defined by the transmitted symbol energy normalized by the noise (i.e., $E_{\mathrm{s}}/N_0$).
The finite-alphabet set used for perturbation is defined by: $\Re(\mathcal{A})=\Im(\mathcal{A})=\{-1, 0, 1\}$, which has shown to be large enough for the performance optimality. 

\begin{figure}
\centering{
\subfigure[$J=8$]{
\includegraphics[scale=0.31]{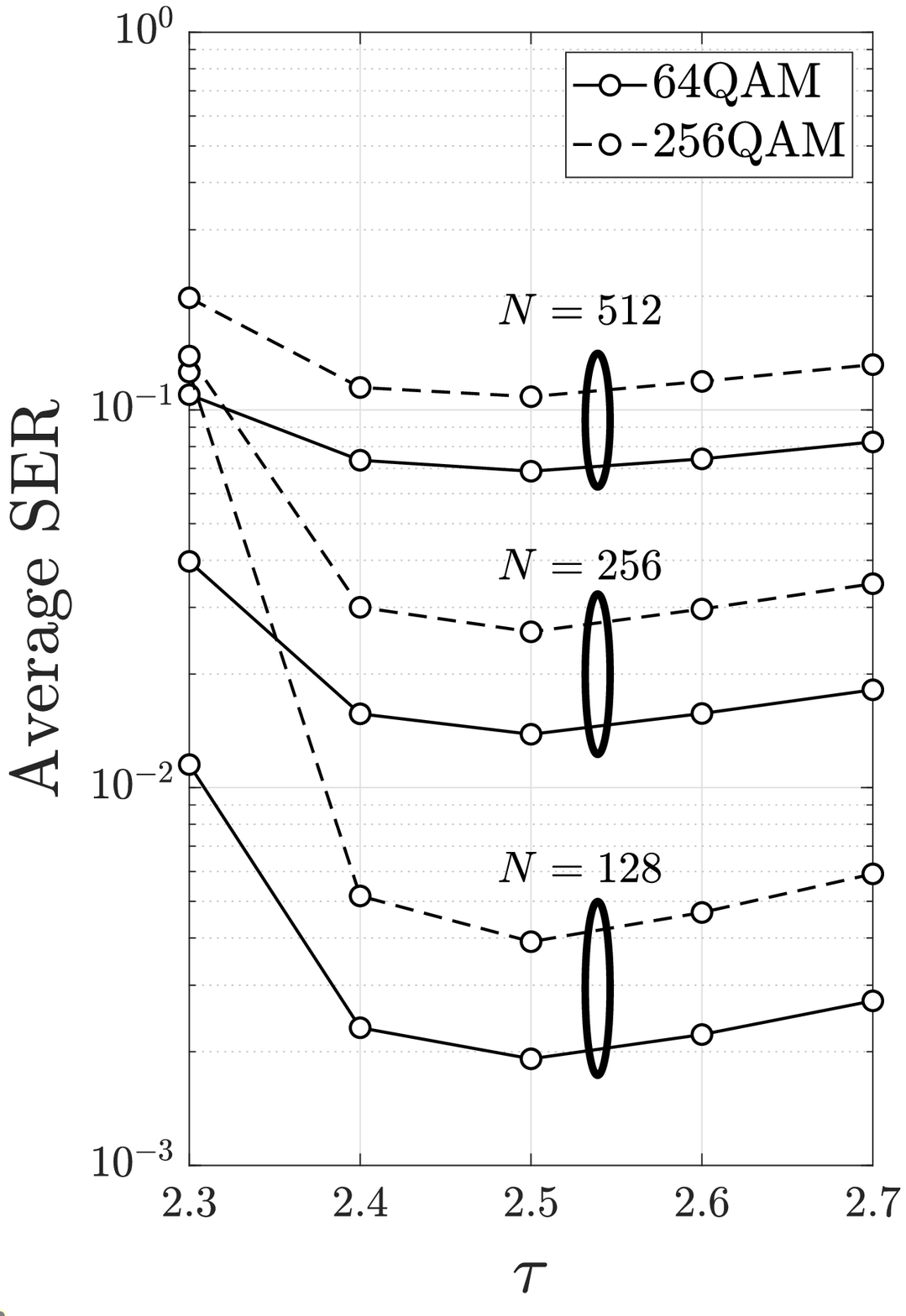}
\label{fig2_a}
\vspace{-1em}
}
\hfil
\subfigure[$J=16$]{
\includegraphics[scale=0.31]{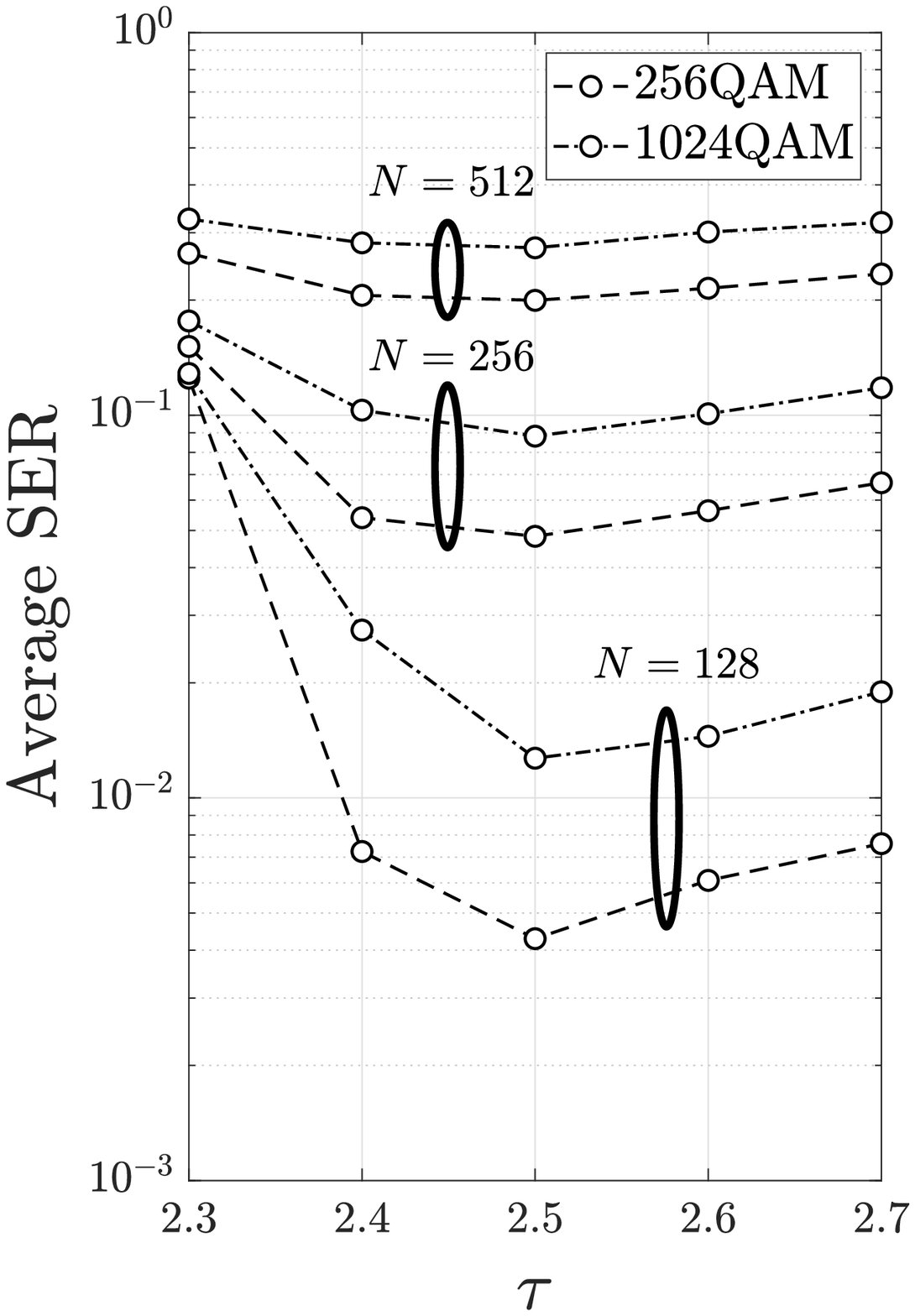}
\label{fig2_b}}
\vspace{-1em}
}
\caption{The average SER as a function of $\tau$. The size of MIMO varies as $N=M=128, 256, 512$; modulations are $64$QAM, $256$QAM and $1024$QAM ($J=8$ or $16$).}
\label{fig2}
\vspace{-1em}
\end{figure} 
The baselines for performance comparison mainly include:
\begin{itemize}
\item Full-dimensional VP: this is the optimum VP precoding. However, due to its exponentially growing complexity with respect to the size of MIMO, full-dimensional VP can only be implemented for small MIMO such as $8$-by-$8$ system.
\item ZF precoding: this is the interference-rejection precoding technique, which is near-optimum for highly over-determined MIMO system but very sub-optimum for large symmetric MIMO. 
\item DKVP ($K=2$): this approach is close to optimum for small MIMO and becomes sub-optimum for large MIMO (see \cite{7397882}).
\item AWGN: this is the case often used as the lower bound for MIMO precoding or signal detection techniques.
\end{itemize}
There exist other approaches such as sphere encoding and lattice reduction and so forth (e.g., \cite{1413598,1369614}). 
As already shown in \cite{7397882}, sphere encoding is of exponential-order of complexity and other sub-optimum approaches are also of very high complexity (at least at the fourth or fifth order of the MIMO size). 
Moreover, they cannot offer a better complexity-performance tradeoff than D2VP.

After all, our computer simulations are divided into three experiments as follows:
\begin{table}[b]
\center
\caption{SNR setup for various modulation sizes and $J$}\label{t01}
\begin{tabular}{|c|c|c|c|}
\hline
&$64$QAM&$256$QAM &$1024$QAM\\
\hline $J=8$& $45$ dB & $50$ dB& --\\
\hline $J=16$& -- & $45$ dB& $50$ dB\\
\hline
\end{tabular}
\end{table}
\subsubsection*{Experiment 1}
Due to the use of widely linear transform, the optimum value of $\tau$ could be different from the optimum value of $\alpha$. 
Moreover, as already discussed in \cite{1413598}, it is very hard to find the optimum of $\tau$ in a mathematical form. 
Therefore, the objective of this experiment is to study the optimum configuration of the parameter ($\tau$) in \eqref{eqn10} through extensive computer simulations.

Fig. \ref{fig2} illustrates the average SER as a function of $\tau$ for various MIMO sizes and higher-order modulations. 
Various system configurations have different SNRs, which are provided in Table \ref{t01}. 
This is to ensure that the average-SNR is mostly located within the range of $(10^{-2}, 10^{-1})$, which is the range of interest for uncoded modulations.

From the illustrated simulation results, it is easy to observe that the minimum SER is achieved for the case of $\tau=2.5$, and this result holds for all the illustrated cases. 
Therefore, we will use $\tau=2.5$ for all the rest of simulations in {\it Experiment 2 \& 3}.

\subsubsection*{Experiment 2}
\begin{figure}[t]
\centering
\includegraphics[scale=0.32]{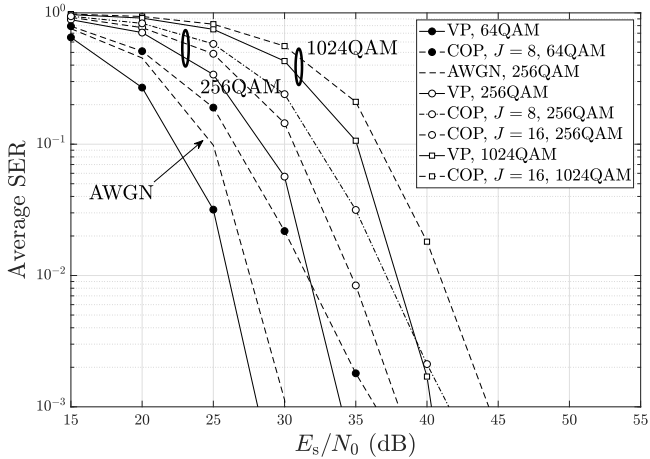}
\caption{Performance comparison between COP and full-dimensional VP in small MIMO ($M=N=8$).}
\vspace{-1em}
\label{fig3}
\end{figure}
\begin{figure}[t]
	\centering
	\includegraphics[scale=0.32]{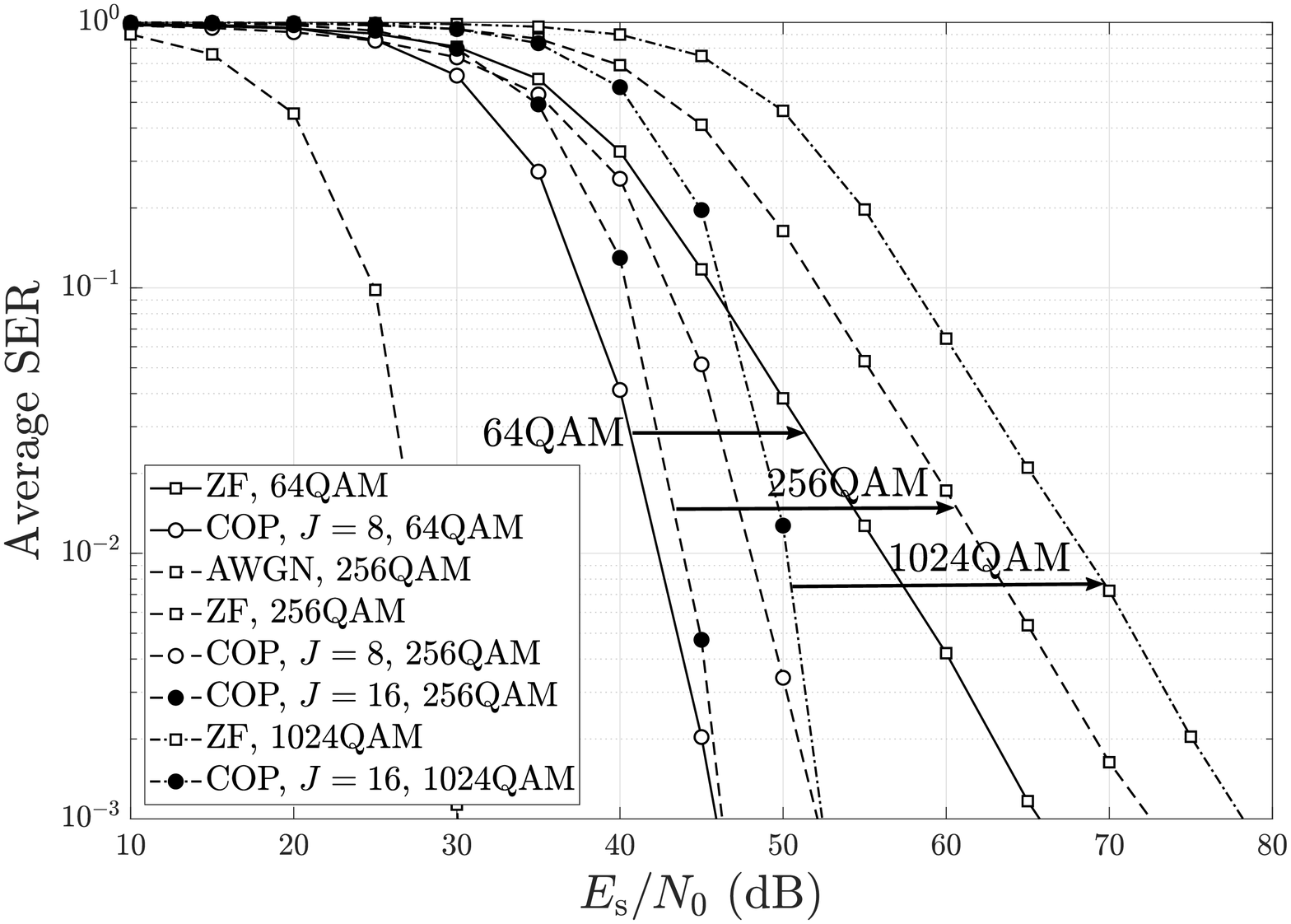}
	\caption{Performance comparison between COP and ZF in large MIMO ($M=N=128$).}
	\vspace{-1em}
	\label{fig4}
\end{figure}

\begin{figure}[t]
	\centering
	\includegraphics[scale=0.32]{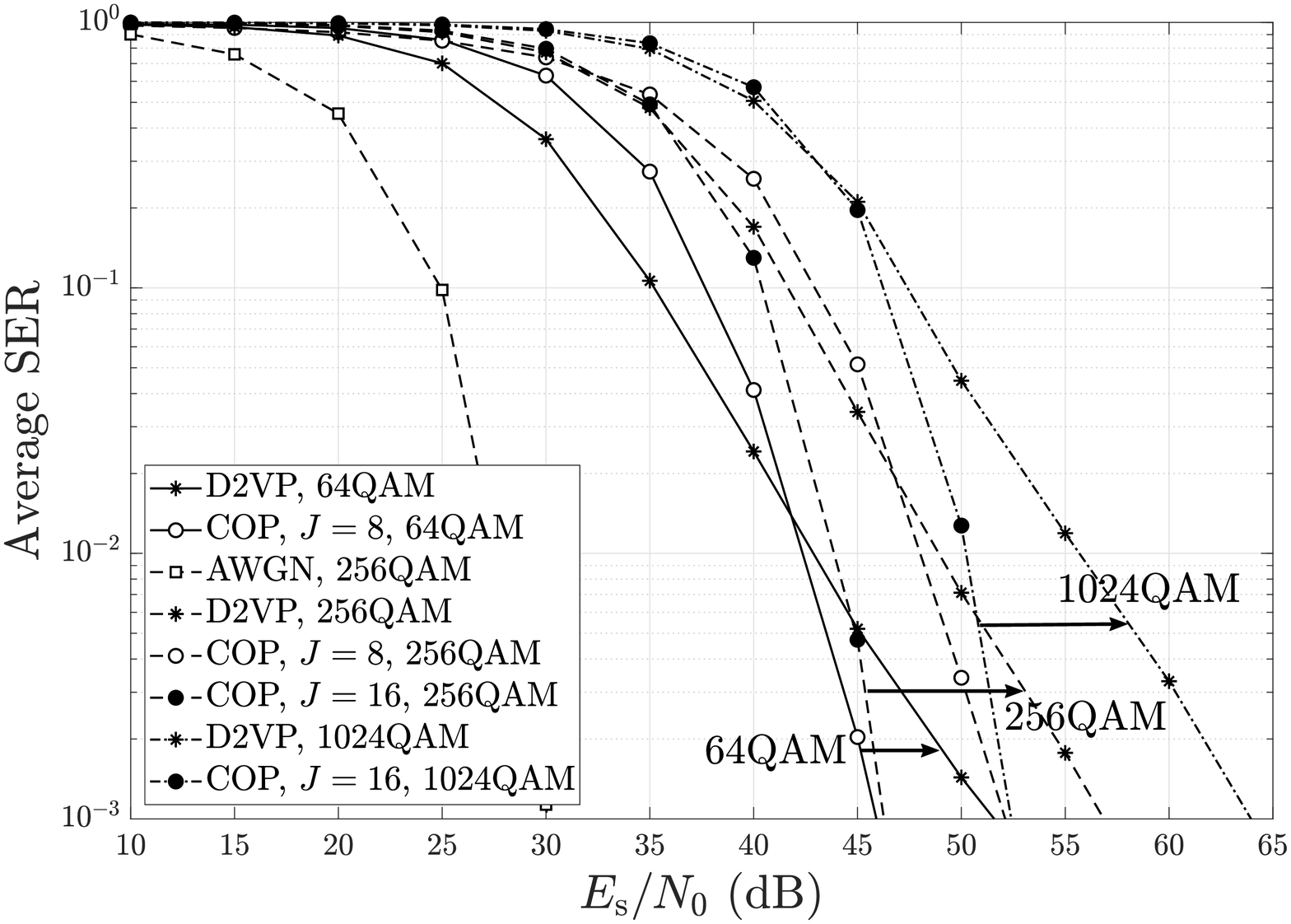}
	\caption{Performance comparison between COP and D2VP in large MIMO ($M=N=128$).}
	\vspace{-1em}
	\label{fig5}
\end{figure}

\begin{figure}[t]
	\centering
	\includegraphics[scale=0.32]{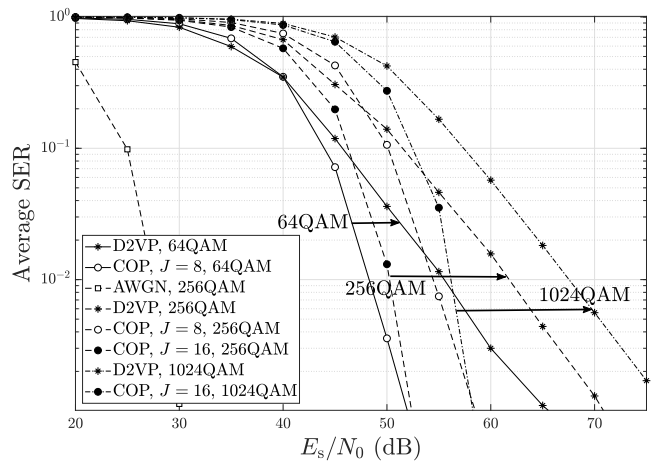}
	\caption{Performance comparison between COP and D2VP in large MIMO ($M=N=512$).}
	\vspace{-1em}
	\label{fig6}
\end{figure}
The objective of this experiment is to conduct performance comparison between COP and full-dimensional VP.
Due to the poor scalability of VP, this experiment can only be conducted for small MIMO ($M=N=8$).
The simulation results of this experiment is plotted in Fig. \ref{fig3}.
Let us start from the case of $256$QAM, where we also show the average SER for the AWGN channel.
It is observed that the full-dimensional VP shows its performance by around $4$ dB away from the AWGN case when SER$=10^{-2}$.
COP ($J=16$) moves further away from the AWGN case by around $3$ dB.
When the complexity of COP is reduced by employing $J=8$, a further $2$ dB performance penalty can be observed.
It is perhaps worth highlighting that the complexity of COP is reduced from $\mathcal{O}(2^{16})$ to $\mathcal{O}(2^8)$. 
This is a significant gain of computational efficiency.

When the modulation order increases to $1024$QAM, the performance gap between COP and full-dimensional VP remains to be around $3$ dB.
For lower range of SER (such as $10^{-1}$), the performance gap reduces to around $1.5$ dB.
It implies that the full-dimensional VP enjoys more spatial diversity gain than the COP technique. 
The reason of this phenomenon has been discussed in the optimality study (see Section \ref{SecIIIc}).
For the comprehensiveness of computer simulations, we also plot the results of $64$QAM. 
For the case of $J=8$, it is observed that the performance gap between COP and the full-dimensional VP is around $5$ dB. 
This shows almost no difference from what we have observed from $256$QAM. 
After all, it can be concluded that COP is a sub-optimum NLP technique with scalable complexity in the constellation domain.
The performance optimality of COP is independent of the modulation order but closely related to the scaling factor ($J$).

\subsubsection*{Experiment 3}
The objective of this experiment is to conduct performance comparison between COP and other precoding techniques approaches for large MIMO systems.

Fig. \ref{fig4} provides the performance comparison between the ZF precoding technique and COP for the MIMO size of  $M=N=128$.
For the case of $J=16$, it is observed that COP outperforms the ZF technique by around $18$ dB for both $256$QAM and $1024$QAM. 
When the scaling factor $J$ is reduced to $8$, the difference between COP and ZF reduces to around $6$ dB (see corresponding curves for $64$QAM and $256$QAM).
Nevertheless, the performance improvement is significant even in this lower complexity version. 
In addition, we also examined the performance gap between COP and the AWGN case (take $256$QAM as an example).
The difference is around $14$ dB when using the scaling factor $J=16$.
This difference does not really mean that COP is too sub-optimum for large MIMO. 
We recall that the full-dimensional VP shows around $4$ dB performance difference with AWGN in the case of $8$-by-$8$ MIMO, and the difference becomes $7\sim 9$ dB when the MIMO size grows to $12$-by-$12$ (see \cite{1413598}). 
Therefore, COP is actually a good sub-optimum NLP technique for large MIMO.

In Fig. \ref{fig5}, we compare the performance between COP and the D2VP approach for the MIMO size of $M=N=128$.
In this context, D2VP is a competitive candidate, whose computational complexity is around $\mathcal{O}(2^{17})$.
For the case of $J=16$ where the complexity of COP is around $\mathcal{O}(2^{16})$, COP outperforms D2VP by around $8$ dB.
When we reduce the scaling factor to $J=8$, the performance gap reduces to $4$ dB whilst the complexity of COP also reduces to $\mathcal{O}(2^8)$.
It is worth noting that COP slightly underperforms D2VP at lower SNRs such as $E_{\mathrm{s}}/N_0<40$ dB.
This is because the performance improvement of COP is mainly from the spatial diversity gain, which becomes significant only at higher SNRs.

When the MIMO size increases to $M=N=512$, Fig. \ref{fig6} shows that D2VP has got significant performance degradation (around $13$ dB).
This is because D2VP conducts perturbation only on two optimally chosen symbols, which is however not enough to maintain the performance optimality. 
It can also be observed that COP only has around $5\sim 6$ dB performance degradation in comparison with the case of $M=N=128$.
As a consequence, COP outperforms D2VP by an additional $6\sim7$ dB gain. 
Finally, it is worth highlighting that the complexity of COP is not affected by the size of MIMO, as it is only related to the scaling factor $J$.
However, the complexity of D2VP increases to $\mathcal{O}(2^{21})$, which is already too large for practical implementation. 

\section{Conclusion}
In this paper, we have introduced a novel NLP technique, namely COP, to tackle the scalability problem inherent in current NLP techniques. 
Unlike conventional techniques that apply perturbation in the symbol domain, COP applies perturbation in the constellation domain. 
With the constellation-domain perturbation, it has been shown that COP has its computational complexity independent of the size of MIMO networks. 
Instead, the complexity is made only related to the modulation order (or the constellation size). 
More appealingly, the complexity of COP is scalable through widely linear transform.
In terms of the performance, COP has been mathematically proved to be a good sub-optimum NLP technique in comparison with the full-dimensional VP. 
The latter is however not scalable to the size of MIMO networks. 
Computer simulations have been conducted to evaluate the performance of COP in various small and large MIMO cases.
It has been demonstrated that COP generally outperforms other sub-optimum VP approaches (mainly D2VP) for large MIMO cases.
More specifically, it can offer $5\sim 20$ dB performance gain with much lower computational complexity in various configurations of MIMO size and modulation. 

\section*{Acknowledgement}
This work was partially funded by the 5G Innovation Centre and the 6G Innovation Centre.

\balance

\ifCLASSOPTIONcaptionsoff
\newpage
\fi

\bibliographystyle{IEEEtran}
\bibliography{Bib_URLLC,Bib_Else,Bib_Precoding}		

\begin{thebibliography}{10}
\providecommand{\url}[1]{#1}
\csname url@samestyle\endcsname
\providecommand{\newblock}{\relax}
\providecommand{\bibinfo}[2]{#2}
\providecommand{\BIBentrySTDinterwordspacing}{\spaceskip=0pt\relax}
\providecommand{\BIBentryALTinterwordstretchfactor}{4}
\providecommand{\BIBentryALTinterwordspacing}{\spaceskip=\fontdimen2\font plus
\BIBentryALTinterwordstretchfactor\fontdimen3\font minus
  \fontdimen4\font\relax}
\providecommand{\BIBforeignlanguage}[2]{{%
\expandafter\ifx\csname l@#1\endcsname\relax
\typeout{** WARNING: IEEEtran.bst: No hyphenation pattern has been}%
\typeout{** loaded for the language `#1'. Using the pattern for}%
\typeout{** the default language instead.}%
\else
\language=\csname l@#1\endcsname
\fi
#2}}
\providecommand{\BIBdecl}{\relax}
\BIBdecl

\bibitem{4599181}
A.~Wiesel, Y.~C. Eldar, and S.~Shamai, ``Zero-forcing precoding and generalized
  inverses,'' \emph{IEEE Trans. Signal Process.}, vol.~56, no.~9, pp.
  4409--4418, Aug. 2008.

\bibitem{1413598}
B.~Hochwald, C.~Peel, and A.~Swindlehurst, ``A vector-perturbation technique
  for near-capacity multiantenna multiuser communication-{Part} {II}:
  Perturbation,'' \emph{IEEE Trans. Commun.}, vol.~53, no.~3, pp. 537--544,
  Apr. 2005.

\bibitem{6151847}
C.~Masouros, M.~Sellathurai, and T.~Ratnarajah, ``Interference optimization for
  transmit power reduction in {Tomlinson-Harashima} precoded {MIMO}
  downlinks,'' \emph{IEEE Trans. Signal Process.}, vol.~60, no.~5, pp.
  2470--2481, Feb. 2012.

\bibitem{5595728}
T.~L. Marzetta, ``Noncooperative cellular wireless with unlimited numbers of
  base station antennas,'' \emph{IEEE Trans. Wireless Commun.}, vol.~9, no.~11,
  pp. 3590--3600, Oct. 2010.

\bibitem{8354786}
J.~C. De~Luna~Ducoing, Y.~Ma, N.~Yi, and R.~Tafazolli, ``A real–complex
  hybrid modulation approach for scaling up multiuser {MIMO} detection,''
  \emph{IEEE Trans. Commun.}, vol.~66, no.~9, pp. 3916--3929, May 2018.

\bibitem{Rusek6375940}
F.~{Rusek}, D.~{Persson}, B.~K. {Lau}, E.~G. {Larsson}, T.~L. {Marzetta},
  O.~{Edfors}, and F.~{Tufvesson}, ``Scaling up {MIMO}: Opportunities and
  challenges with very large arrays,'' \emph{IEEE Signal Process. Mag.},
  vol.~30, no.~1, pp. 40--60, Dec. 2012.

\bibitem{1350238}
L.~Bruhl, C.~Degen, W.~Keusgen, B.~Rembold, and C.~Walke, ``Investigation of
  front-end requirements for {MIMO}-systems using downlink pre-distortion,'' in
  \emph{2003 5th Euro. Personal Mob. Commun. Conf. (Conf. Publ. No. 492)},
  2003, pp. 472--476.

\bibitem{BJORNSON20193n}
E.~{Bj\"ornson}, L.~{Sanguinetti}, H.~{Wymeersch}, J.~{Hoydis}, and T.~L.
  {Marzetta}, ``Massive {MIMO} is a reality–{What} is next? {Five} promising
  research directions for antenna arrays,'' \emph{Digit. Signal Process.},
  vol.~94, pp. 3--20, Nov. 2019.

\bibitem{Wang2022}
\BIBentryALTinterwordspacing
J.~Wang, Y.~Ma, N.~Yi, R.~Tafazolli, and F.~Wang, ``Network-{ELAA} beamforming
  and coverage analysis for {eMBB/URLLC} in spatially non-stationary {Rician}
  channels,'' {IEEE} Int. Conf. Commun. (ICC) 2022. [Online]. Available:
  \url{https://arxiv.org/abs/2201.07875}
\BIBentrySTDinterwordspacing

\bibitem{9685536}
J.~Liu, Y.~Ma, J.~Wang, N.~Yi, R.~Tafazolli, S.~Xue, and F.~Wang, ``A
  non-stationary channel model with correlated {NLoS/LoS} states for
  {ELAA-mMIMO},'' in \emph{Proc. IEEE Global Commun. Conf. (GLOBECOM)}, Dec.
  2021, pp. 1--6.

\bibitem{7397882}
Y.~Ma, A.~Yamani, N.~Yi, and R.~Tafazolli, ``Low-complexity {MU-MIMO} nonlinear
  precoding using degree-2 sparse vector perturbation,'' \emph{IEEE J. Sel.
  Areas Commun.}, vol.~34, no.~3, pp. 497--509, Feb. 2016.

\bibitem{7331280}
J.~C. De~Luna~Ducoing, N.~Yi, Y.~Ma, and R.~Tafazolli, ``Using real
  constellations in fully- and over-loaded large {MU-MIMO} systems with simple
  detection,'' \emph{IEEE Wireless Commun. Lett.}, vol.~5, no.~1, pp. 92--95,
  Feb. 2016.

\bibitem{7882699}
W.~Zhang, R.~C. de~Lamare, C.~Pan, M.~Chen, J.~Dai, B.~Wu, and X.~Bao, ``Widely
  linear precoding for large-scale {MIMO} with {IQI}: Algorithms and
  performance analysis,'' \emph{IEEE Trans. Wireless Commun.}, vol.~16, no.~5,
  pp. 3298--3312, May 2017.

\bibitem{1369614}
C.~Windpassinger, R.~Fischer, and J.~Huber, ``Lattice-reduction-aided broadcast
  precoding,'' \emph{IEEE Trans. Commun.}, vol.~52, no.~12, pp. 2057--2060,
  Dec. 2004.

\end{thebibliography}
\end{document}